\documentclass[aps,prd,showpacs,floatfix,superscriptaddress,preprintnumbers,nofootinbib]{revtex4}
\usepackage{latexsym}
\usepackage{amssymb}
\usepackage{amsmath}

\begin{document}

\title{On the theory of spherically symmetric thin shells in conformal gravity}
\author{Victor Berezin}\thanks{e-mail: berezin@inr.ac.ru}
\affiliation{Institute for Nuclear Research of the Russian Academy of
	Sciences \\ 60th October Anniversary Prospect 7a, 117312 Moscow, Russia}
\author{Vyacheslav Dokuchaev}\thanks{e-mail: dokuchaev@inr.ac.ru}
\affiliation{Institute for Nuclear Research of the Russian Academy of
		Sciences \\ 60th October Anniversary Prospect 7a, 117312 Moscow, Russia}
\affiliation{National Research Nuclear University ``MEPhI'' (Moscow Engineering Physics Institute) \
Kashirskoe shosse 31, Moscow, 115409, Russia}
\author{Yury Eroshenko}\thanks{e-mail: eroshenko@inr.ac.ru}
\affiliation{Institute for Nuclear Research of the Russian Academy of
		Sciences \\ 60th October Anniversary Prospect 7a, 117312 Moscow, Russia}
	
	\date{\today}

\begin{abstract}
The spherically symmetric thin shells are the nearest generalizations of the point-like particles. Moreover, they serve as the simple sources of the gravitational fields both in General Relativity and much more complex quadratic gravity theories. We are interested in the special and physically important case when all the quadratic in curvature tensor (Riemann tensor) and its contractions (Ricci tensor and scalar curvature) terms are present in the form of the square of Weyl tensor. By definition, the energy-momentum tensor of the thin shell is proportional to Dirac delta-function. We constructed the theory of the spherically symmetric thin shells for three types of gravitational theories with the shell: (1) General Relativity; (2) Pure conformal (Weyl) gravity where the gravitational part of the total Lagrangian is just the square of the Weyl tensor; (3) Weyl$+$Einstein gravity. The results are compared with these in General Relativity (Israel equations). We considered in details the shells immersed in the vacuum. Some peculiar properties of such shells are found. In particular, for the traceless ($=$ massless) shells it is shown that their dynamics can not be derived from the matching conditions and, thus, is completely arbitrary. On the contrary, in the case of the Weyl$+$Einstein gravity the trajectory of the same type of shell is completely restored even without knowledge of the outside solution.
\end{abstract}

\pacs{04.20.Fy, 04.20.Jb, 04.50.Kd, 04.60.Bc, 04.70.Bw}


\maketitle

\section{Introduction}	

It is well known that the thin shells play an important role in General Relativity by providing us with the rather simple models  of the self-gravitating matter sources, which, in turn, develop our physical intuition about the very complex structure of the whole theory. In the case of the spherical symmetry the thin shells, in addition, are the simplest possible generalization of the point particles.

What is the thin shell? By definition, in the $(d+1)$--dimensional manifold, this is a $d$-dimensional hypersurface $\Sigma$, which divides the manifold into two parts and where the energy-momentum tensor is concentrated. That is, the energy-momentum tensor $T_{\mu\nu}\big|_\Sigma=S_{\mu\nu}\delta(\Sigma)$, where $S_{\mu\nu}$ is called the surface energy-momentum tensor, and $\delta(\Sigma)$ is the Dirac delta--function. In general, on such a hypersurface there may be a jump $[T_{\mu\nu}]\Theta(\Sigma)$, where $\Theta(\Sigma)$ is the Heaviside step function. When $S_{\mu\nu}\neq0$, the hypersurface $\Sigma$ is called singular. Thus, the thin shell is some singular hypersurface, where two parts of the space-time manifold should be matched. In what follows, we will be dealing only with four-dimensional space-time and, accordingly, with three-dimensional thin shells.

The thin shell formalism in General Relativity was elaborated by W. Israel \cite{Isr,Cruz67} in 1966, and since then it was widely used for investigations of cosmological phase transitions (see 
\cite{Lake79}--\cite{Mann08} for more details) and black hole structures (see, e.\,g.,  \cite{Boulware73}--\cite{Berezin14b}).

The matching hypersurface is characterized by its three-dimensional metric and by the extrinsic curvature tensor that shows how this hypersurface is embedded into the four-dimensional space-time. And the Israel equations relate the surface energy-momentum tensor to the jumps in the extrinsic curvature tensor on both sides of the thin shell. Some note is in order here.  Since the space-time is considered continuous on the matching hypersurface $\Sigma$ (otherwise we would deal with the two disconnected manifolds) and the Einstein equations are of the second order in derivatives of the metric tensor, then the first derivatives may have, at most, the jumps, while the second derivatives may exhibit, at most, the delta--function behavior. Therefore, the most singular of the allowed types of the matching hypersurfaces, containing the nonzero energy-momentum tensor, are just the thin shells. As a consequence, the thin shell produces the real space-time singularity, where the Riemann curvature tensor diverges. Thats how the things are in General Relativity. 

Started in 1969, there was a long history and big activity in studying the quantum field theories in curved space-times (see Refs.~\cite{Parker69}--\cite{Zeld77} for more details). Here the most interesting for us is the main result, that the renormalization procedure, inevitable in the quantum field theory, leads to the appearance in the effective action of some additional (to the standard Einstein-Hilbert Lagrangian, linear in the curvature scalar) terms quadratic in the Riemann tensor and its contractions (Ricci tensor and scalar curvature). When the gravitational Lagrangian  is the general combination of these quadratic terms plus the linear term and plus the cosmological constant, it is usually called ``the quadratic gravity''. Some of its special form was used by A. Starobinsky \cite{Alstar} for constructing the first inflationary cosmological scenario. The necessity of considering the quadratic and and higher order terms in the effective gravitational Lagrangian, as was foreseen by A. D. Sakharov \cite{Sakh} in 1967. He even suggested that the gravitation itself is just the tensions and stresses in the vacuum of all other quantum fields. Now this idea is widely known as ``the induced gravity''.

The thin shell formalism for the generic quadratic gravity was constructed in \cite{Frolov}. The main difference from General Relativity is that the equations of motion appeared to be fourth order in the derivatives of the metric tensor (this feature is shared by the so called $f(R)$--theories of gravity, and the exceptions are the Horndeski and Gaus--Bonnet models). It means that if we restrict ourselves with the singularities in the equations of motion not higher than the Dirac's delta function,  then the Riemann curvature tensor should be continuous on the thin shell and (in general) not singular. Therefore, the matching conditions (the analog of Israel equations) are also quite different from that of General Relativity. One of the combinations of the quadratic terms is very peculiar. This is the case of the conformal gravity, when the quadratic part of the gravitational Lagrangian is just the square of the Weyl tensor, which  is the completely traceless part of the Riemann curvature tensor. Note, that in the absence of the linear term it is known under the names ``Weyl geometry'' or ``pure conformal gravity''. The matter is that the Weyl tensor itself (with only one upper index) is invariant under the local conformal transformation (when the whole metric tensor is multiplied by some function called ``the conformal factor''). The resulting Bach tensor, which appeared in the gravitational equations of motion, is multiplied by some power of this conformal factor, without any of its derivatives in the equations of motion. But, we are free to choose as a conformal factor any of the components of metric tensor. Thus, we should only require its continuity on the shell, but not the continuity of the derivatives. So, in some cases, the Riemann tensor constructed from the initial metric, may have the delta--function structure on the thin shell.

The above consideration is especially important in the case of the spherical symmetry because we are able to use the radius of the sphere as the conformal factor (see Refs.~\cite{BDE16}--\cite{BDE17} for more details), and it is the radius as function of the proper time on the shell, that determines completely its trajectory in the space-time. 

Recently, J. M. M. Senovilla \cite{Senovilla15} (see also \cite{Senovilla17}) discovered that, in the framework of the theory of distributions, the quadratic gravity (as well as all other modifications of the Einstein gravity leading to the fourth order differential equations) allows not only the thin shell (the delta--function distributions), but also the double layers (the delta-prime distributions). This is a very compelling possibility, though its physical meaning is not yet clear due to the absence of the corresponding counterpart in the energy-momentum tensor for the conventional matter (the absence of particles with negative masses). In what follows we will not consider the double layers at all.

The aim of the present paper is to investigate and compare thin shells in three different gravitational theories: General Relativity (Einstein gravity), pure conformal (Weyl) gravity and more general conformal gravity (Weyl+Einsten). For the sake of simplicity we restrict ourselves to the case of the spherical symmetry.

\section{Preliminaries}
\label{prelim}

In this Section we draw up some needed formulas. Throughout the paper we will be using the $(+,-,-,-)$--signature for the line element
\begin{equation}
ds^2=g_{\mu\nu}dx^\mu dx^\nu \quad (\mu,\nu=0,1,2,3).
\end{equation}
The metric connections 
\begin{equation}
\Gamma_{\mu\nu}^\lambda=\frac{1}{2} g^{\lambda\sigma}(g_{\sigma\mu,\nu}+g_{\sigma\nu,\mu}-
g_{\mu\nu,\sigma}),
\end{equation}
where comma denotes the partial derivative. The Riemann curvature tensor is defined as 
\begin{equation}
R^\mu_{\nu\lambda\sigma}=\frac{\partial\Gamma^\mu_{\nu\sigma}}{\partial x^\lambda}
-\frac{\partial\Gamma^\mu_{\nu\lambda}}{\partial x^\sigma}
+\Gamma^\mu_{\varkappa\lambda}\Gamma^\varkappa_{\nu\sigma}
-\Gamma^\mu_{\varkappa\sigma}\Gamma^\varkappa_{\nu\lambda},
\end{equation}
while the Ricci tensor equals
\begin{equation}
R_{\mu\nu}=g^{\lambda\sigma}R_{\lambda\mu\sigma\nu}=R^\lambda_{\mu\lambda\nu},
\end{equation}
and its convolution gives the curvature scalar
\begin{equation}
R=g^{\lambda\sigma}R_{\lambda\sigma}.
\end{equation}
The Einstein tensor
\begin{equation}
G_{\mu\nu}=R_{\mu\nu}-\frac{1}{2}g_{\mu\nu}R.
\end{equation}
The Weyl tensor (completely traceless part of the Riemann tensor) 
\begin{eqnarray}
C_{\mu\nu\lambda\sigma}&=&R_{\mu\nu\lambda\sigma}+\frac{1}{2}(R_{\mu\sigma}g_{\nu\lambda}
+R_{\nu\lambda}g_{\mu\sigma}-R_{\mu\lambda}g_{\nu\sigma}-R_{\nu\sigma}g_{\mu\lambda}) \nonumber \\
&&+\frac{1}{6}R(g_{\mu\nu}g_{\lambda\sigma}-g_{\mu\sigma}g_{\nu\lambda}),  
\end{eqnarray}
The Bach tensor is
\begin{equation}
B_{\mu\nu}=C_{\mu\lambda\nu\sigma}^{\phantom{0000};\sigma;\lambda}
+\frac{1}{2}R^{\lambda\sigma}C_{\mu\lambda\nu\sigma}, 
\end{equation}
\begin{equation}
B^{\lambda}_{\lambda}=0, \quad B_{\mu\nu}=B_{\nu\mu}, \quad B^\lambda_{\mu;\lambda}=0,
\end{equation}
(semicolon ``;'' denotes the covariant derivative with respect to the metric $g_{\mu\nu}$).

Next, we introduce the local conformal transformation by
\begin{equation}
ds^2=\Omega^2(x) d\hat s^2, \quad g_{\mu\nu}=\Omega^2(x) \hat g_{\mu\nu}, \quad 
g^{\mu\nu}=\frac{1}{\Omega^2(x)} \hat g^{\mu\nu},
\end{equation}
where the hat ``$\;\hat{}\;$'' means ``conformally transformed''.

Then, the Einstein tensor is transformed as
\begin{equation}
G_{\mu\nu}=\hat G_{\mu\nu}-\frac{2\Omega_{\mu|\nu}}{\Omega}
+\frac{2\Omega^\lambda_{\phantom{0}|\lambda}}{\Omega}\hat g_{\mu\nu}
+\frac{4\Omega_{\mu}\Omega_{\nu}}{\Omega^2}-\frac{\Omega^\lambda \Omega_\lambda}{\Omega^2}\hat g_{\mu\nu},
\end{equation}
where $\Omega_\mu=\Omega_{,\mu}$, $\Omega^\lambda=\hat g^{\lambda\sigma}\Omega_\sigma$, and the vertical line ``$|$'' denotes the covariant derivative with respect to the transformed metric $\hat g_{\mu\nu}$. The Weyl tensor with only one upper index is invariant under the local conformal transformation 
\begin{equation}
C^\mu_{\nu\lambda\sigma}=\hat C^\mu_{\nu\lambda\sigma},
\end{equation}
so 
\begin{equation}
C^2\sqrt{-g}=C_{\mu\nu\lambda\sigma}C^{\mu\nu\lambda\sigma}\sqrt{-g}=\hat C^2\sqrt{-\hat g},
\end{equation}
and the Bach tensor is transformed as
\begin{equation}
B_{\mu\nu}=\frac{1}{\Omega^2} \hat B_{\mu\nu}.
\end{equation}
If one decompose the total action integral into the gravitational and non-gravitational parts, $S_{\rm tot}=S_{\rm gr}+S_{\rm m}$, then the energy -momentum tensor $T_{\mu\nu}$ and its ``transformed'' counterpart
$\hat T_{\mu\nu}$ are defined through the following variations of $S_{\rm m}$:
\begin{eqnarray}
\delta S_{\rm m}&& \stackrel{def}{=}
\frac{1}{2}\int\! T_{\mu\nu}\sqrt{-g}\delta g^{\mu\nu}dx,  \\
\delta S_{\rm m}&& \stackrel{def}{=} \frac{1}{2}\int\! \hat T_{\mu\nu}\sqrt{-\hat g}\delta \hat g^{\mu\nu}dx,
\end{eqnarray}
Thus, 
\begin{equation}
\hat T_{\mu\nu}=\Omega^2T_{\mu\nu}, \quad \hat T_\mu^\nu=\Omega^4T_\mu^\nu, \quad \hat T^{\mu\nu}=\Omega^6T^{\mu\nu}.
\end{equation}

Let us now come to the spherical symmetry. The spherical symmetric line element has the form
\begin{equation}
ds^2=g_{\mu\nu}dx^\mu dx^\nu = \gamma_{ik}dx^i dx^k - r^2(x)(d\theta^2+\sin^2\theta d\phi^2),
\quad (i,k=0,1),
\label{ds2}
\end{equation}
where $\theta$ and $\phi$ are the usual spherical angles, $r(x)$ is the radius of the sphere (with the area $4\pi r^2$), and $\gamma_{ik}$ is the metric tensor of the two-dimensional space-time. It appears convenient to make the conformal transformation and choose the radius $r(x)$ as the conformal factor, $\Omega(x)=r(x)$. Then,
\begin{equation}
d\hat s^2 = \tilde\gamma_{ik}dx^idx^k - (d\theta^2+\sin^2\theta d\phi^2).
\label{dhats2}
\end{equation}
We can make the $(2+2)$--decomposition and reduce the problem, actually, to the two-dimensional one. Namely
\begin{eqnarray}
G_{ik}&=&-\frac{2r_{i||k}}{r}+\frac{4r_ir_k}{r^2} \label{Gik}
+\left(1+\frac{2r^p_{\phantom{0}||p}}{r} -\frac{r^p r_p}{r^2}\right)\tilde\gamma_{ik},  \\
G&=&G^\lambda_\lambda=-R=-\frac{1}{r^2}\left(-\hat R+\frac{6r^p_{\phantom{0}||p}}{r}\right), \quad
\hat R=\tilde R-2,
\label{G}
\end{eqnarray}
where $\tilde R$ is the two-dimensional curvature scalar constracted of the metric $\tilde\gamma_{ik}$, and the double vertical line ``$||$''  denotes the covariant derivative with respect to this very metric.

The transformed Bach tensor becomes 
\begin{equation}
B_{\mu\nu}=\frac{1}{r^2}\hat B_{ik}, \quad 
 \hat B_{ik}=-\frac{1}{6}\left(\tilde R^{||p}_{||p}\tilde\gamma_{ik}-\tilde R_{||ik}+  
\frac{\tilde R^2-4}{4}\tilde\gamma_{ik}\right).
\end{equation}
We do not need the separate expressions for the remaining components, since due to the spherical symmetry $\hat B_2^2=\hat B_3^3$, and the tracelessness of the Bach tensor, $B_\mu^\mu=0$, yields
\begin{equation}
B_2^2=\hat B_3^3=-\frac{1}{2}\hat B_p^p.
\end{equation}
 
Coming to the point, we should describe our spherically symmetric thin shell, $\Sigma$. It is simply a sphere which radius, $\rho$, evolving in time. The corresponding line element equals 
\begin{eqnarray}
ds^2|_\Sigma&=&d\tau^2-\rho^2(\tau) (d\theta^2+\sin^2\theta d\phi^2) \nonumber  \\
&=&\rho^2(\tau)\left(d\tilde\tau^2- (d\theta^2+\sin^2\theta d\phi^2)\right).
\end{eqnarray}
Hence, $\tau$ is the proper time of the observer sitting on the shell (we may call it ``the genuine proper time''), while $\tilde\tau$ is its conformal counterpart (we may call it ``the conformal proper time''). The simplest way to proceed is to use the two-dimensional part of the conformally transformed metric,  
$d\hat s^2 = \tilde\gamma_{ik}dx^idx^k$, $(i,k=0,1)$. Evidently, it is always possible to introduce the so called Gauss normal coordinate system, associated with the given thin shell. The line element has the following form, valid both on the shell as well as just outside it,
\begin{equation}
d\hat s_2^2=\tilde\gamma_{00}(n,\tilde\tau)d\tilde\tau^2-d\tilde n^2,
\label{dhats22}
\end{equation}
where the normal coordinate $n$ runs from inward $n<0$ to outward $n>0$, and the equation of our thin shell is simply $n=0$. To match the above line element with that  on the shell, we impose the normalization conditions 
\begin{equation}
\tilde\gamma_{00}(0,\tilde\tau)=1.
\end{equation}

The extrinsic curvature tensor, describing the embedding of the shell (in our case it is just the worldline $\tilde n=0$) into the ambient (two-dimensional) space-time, 
\begin{equation}
\hat K_{ij}= -\frac{1}{2}\frac{\partial\hat g_{ij}}{\partial n},
\end{equation}
now consists of only one component
\begin{equation}
\hat K_{00}\stackrel{def}{=} -\frac{1}{2}\frac{\partial\tilde\gamma_{00}}{\partial n}=\tilde K_{00},
\quad \tilde K=\tilde K^0_0= -\frac{1}{2}\frac{\partial\log\tilde\gamma_{00}}{\partial n},
\end{equation}
and the two-dimensional curvature scalar equals
\begin{equation}
\tilde R=-2\tilde K_n+2\tilde K^2.
\end{equation}

With the use of Gauss normal coordinates the transformed energy-momentum tensor can be easily written as 
\begin{equation}
\hat T_{\mu\nu} \stackrel{def}{=} \hat S_{\mu\nu}\delta(n)
+[\hat T_{\mu\nu}]\Theta(n)+\hat T_{\mu\nu}^{(-)},
\end{equation}
where $\hat S_{\mu\nu}$ is the surface energy-momentum tensor  of the shell, $\delta(n)$ is the Dirac's delta--function, $\Theta(n)$ is the Heaviside step--function.
\begin{equation}
\Theta(n)=\left\{
\begin{array}{rl}	
1, & \mbox{if } n>0 \;\; (+), \\
0, & \mbox{if } n<0 \;\; (-). 
\end{array}	
\right. 
\end{equation}
We will need the following properties of $\Theta$--function:
\begin{equation}
\Theta^2=\Theta, \quad \Theta'(n)=\delta(n),
\end{equation}
and $[\ldots]=$ denotes the ``jump'' across the shell in the outward normal  direction. Thus,
\begin{equation}
[\hat T_{\mu\nu}]=[\hat T_{\mu\nu}^{(+)}-\hat T_{\mu\nu}^{(-)}].
\end{equation}

\section{Thin shells in gravity theories}

In the remaining Section we will derive the matching conditions for the spherically symmetric space-time containing thin shells for three different theories: Einstein gravity ($=$ General Relativity), Weyl gravity ($=$ pure conformal gravity) and Einstein$+$Weyl gravity.  In order to be able to compare the results we consider only one type of the spherically symmetric shells, namely, those with traceless surface energy-momentum tensor, immersed into the vacuum. By the way, the traceless spherical shells can be realized as the shells with massless orbiting constituents \cite{Berezin01}. In addition, we suppose that such a shell is the innermost one, so there are no other gravitating sources inside it. To be more precise, the vacuum inside will be that with zero Weyl tensor, corresponding to $\tilde R=2$.

\subsection{Thin shells in Einstein gravity}

For the sake of completeness we start from the very beginning, i.\,e., from the total action integral, $S_{\rm tot}=S_{\rm gr}+S_{\rm m}$, with the gravitational part $S_{\rm gr}$ being the Einstein-Hilbert action including the cosmological term $\Lambda$. Namely,
\begin{equation}
S_{\rm tot}=\frac{1}{16\pi G}\int(R-2\Lambda)\sqrt{-g}\,dx+S_{\rm m}.
\end{equation}
The equation of motion, of course, are (without hats)
\begin{equation}
\frac{1}{8\pi G}(G_{\mu\nu}-\Lambda g_{\mu\nu})=T_{\mu\nu}.
\end{equation}
The chosen spherically symmetric metric with the radius $r$ as the conformal factor may be written in the form (\ref{dhats2}). Respectively, the two dimensional part of this metric with the use of Gauss normal coordinates takes the form  (\ref{dhats22}). The resulting Einstein equations (\ref{Gik}) now  can be written in the following way
\begin{eqnarray}
(00) \; &\Rightarrow& \; -\frac{2r_{,nn}}{r}\tilde\gamma_{00}+\frac{r_{,n}^2}{r}\tilde\gamma_{00}
+\frac{3\dot r^2}{r^2}+1=\frac{8\pi G}{r^2}\hat T_{00};  \\
(0n) \; &\Rightarrow& \; -\frac{2}{r}(\dot r_{,n}+\tilde K\dot r)+\frac{4\dot rr_{,n}}{r^2}
=\frac{8\pi G}{r^2}\hat T_{0n};   \\
(nn) \; &\Rightarrow& \; \frac{3}{r^2}r_{,n}^2-1+\frac{2}{\tilde\gamma_{00}}\frac{\ddot r}{r}
+\frac{\dot{\tilde{\gamma}}_{00}}{\tilde\gamma_{00}^2}\frac{\dot r}{r}
+\frac{\dot r^2}{\tilde\gamma_{00}r^2}-\frac{2\tilde Kr_{,n}}{r}
=\frac{8\pi G}{r^2}\hat T_{nn}-\Lambda r^2;   \\
(Tr)  &\Rightarrow&1\!+\!\tilde K_{,n}\!-\!\tilde K^2\!+\!\frac{3}{r}\left(\frac{\ddot r}{\tilde\gamma_{00}}
-\frac{1}{2}\frac{\dot{\tilde{\gamma}}_{00}}{\tilde\gamma_{00}^2}\dot r
\!+\!\tilde Kr_{,n}\!-\!r_{,nn}\!\right)\!=\!\frac{4\pi G}{r^2}Tr\hat T\!+\!2\Lambda r^2\!.
\end{eqnarray}
Since $\hat T_{\mu\nu}=\hat S_{\mu\nu}\delta(n)
+[\hat T_{\mu\nu}]\Theta(n)+\hat T_{\mu\nu}^{(-)}$, we should write down the same decompositions for $r$, $\tilde\gamma_{00}$ and $K$, insert them into equations and then equate on left and right sides separately the terms in front of $\delta(n)$ and $\Theta(n)$, remembering that $[r]=0$ and $\tilde\gamma_{00}=0$. So,
\begin{equation}
\left\{
\begin{array}{ll}	
r_{,n}=[r_{,n}]\Theta(n)+r_{,n}^{(-)} &  \\
r_{,nn}=[r_{,n}]\delta(n)+[r_{,nn}]\Theta(n)+r_{,nn}^{(-)} & 
\end{array}	\right.
\end{equation}
\begin{equation}
\left\{
\begin{array}{ll}	
\tilde K=[\tilde K]\Theta(n)+\tilde K^{(-)} &  \\
\tilde K_{,n}=[\tilde K]\delta(n)+[\tilde K_{,n}]\Theta(n)+\tilde K_{,n}^{(-)} & 
\end{array}	
\right.
\end{equation}
Note that the matching conditions are, in fact, the Einstein equations on $\Sigma$ ($=$ Israel equations). They are 
\begin{equation}
\!\!\!\!\!\!\!\!\!\!\!\!\!\!\!\!\!\!\!\!\!\!\!(00) \ \Rightarrow \ \left\{
\begin{array}{ll}	
-\frac{2}{r}[r_{,n}]=\frac{8\pi G}{r^2}\hat S_{00}; &  \\
-\frac{1}{r^2}[r_{,n}^2]-\frac{2}{r}[r_{,nn}]=\frac{8\pi G}{r^2}[\hat T_{00}]. &  
\end{array}	\right.
\end{equation}
\begin{equation}
(0n) \ \Rightarrow \ \left\{
\begin{array}{ll}	
\hat S_{0n}=0; &  \\
-\frac{2}{r}[r_{,n}]-\frac{2\dot r}{r}[\tilde K]+\frac{4 \dot r}{r^2}[r_{,n}]=\frac{8\pi G}{r^2}\hat [T_{0n}]. &  
\end{array}	\right.
\end{equation}
\begin{equation}
\!\!\!\!\!\!\!\!\!\!\!\!\!\!\!\!\!\!\!\!(nn) \ \Rightarrow \ \left\{
\begin{array}{ll}	
\hat S_{nn}=0; &  \\
\frac{3}{r^2}[r_{,n}]-\frac{2}{r}[\tilde Kr_{,n}]=\frac{8\pi G}{r^2}[\hat T_{nn}]. &  
\end{array}	\right.
\end{equation}
\begin{equation}
(Tr) \ \Rightarrow \ \left\{
\begin{array}{ll}	
[\tilde K]-\frac{3}{r}[r_{,n}]=\frac{4\pi G}{r^2}(Tr\hat S); &  \\
\![\tilde K_{,n}]-[\tilde K^2]+\frac{3}{r}([\tilde Kr_{,n}]-[r_{,nn}])=\frac{4\pi G}{r^2}[Tr\hat T]. &  
\end{array}	\right.
\end{equation}
For simplicity we do not consider here the jump in the cosmological constant across the shell. One can get rid of $\dot r_n$ and $[\tilde K]$ in the second, $(0n)$, equation, and rewrite the set of the shell equations in the following way
\begin{equation}
\left\{
\begin{array}{ll}	
\dot{\hat{S}}_0^0-\frac{\dot\rho}{\rho}(\hat{S}_0^0+2\hat{S}_2^2)+[\hat T_0^n]=0; &  \\
-[r_{,n}]=\frac{4\pi G}{\rho^2}\hat S_0^0; &  \\
-[\tilde K]=\frac{8\pi G}{\rho^2}(\hat S_0^0-\hat S_2^2),
\end{array}	\right.
\end{equation}
where  we introduced $\rho=r(\tilde\tau,0)=r\big|_\Sigma$ --- the radius of the shell as a function of the conformal proper time $\tilde\tau$. In what follows we do not need all other equations since they are not relevant to the determining the shell's trajectory we are interested in. The first equation is nothing more but the shell's surface energy-momentum tensor conservation equation, the jump  $[\hat T_0^n]$ being show what amount of the bulk energy flux is absorbed or radiated by the shell. Note that if  $[\tilde K]\big|_\Sigma\neq0$, i.\,e., $\hat S^0_0\neq \hat S^2_2$, then the curvature scalar $\tilde R$ of the two-dimensional conformal space-time is singular on the shell. To complete the investigation we should know what kind of the shell is under consideration and what are its environment on both sides. As we already mentioned we will consider  only traceless shells (i.\,e.,  $\hat S^0_0+2 \hat S^2_2=0$) and the vacuum space-times on both sides. Thus the first equation now reads simply 
\begin{equation}
\dot{\hat{S}}_0^0=0  \quad  \Rightarrow  \quad \hat{S}_0^0=S_0=const.
\end{equation}
So, our shell equations become
\begin{equation}
\left\{
\begin{array}{ll}	
-[r_{,n}]=\frac{4\pi G}{\rho^2}S_0; &  \\
-[\tilde K]=\frac{12\pi G}{\rho^2}S_0.
\end{array}	\right.
\end{equation}
Let us start from the first of these equations and introduce the invariant
\begin{equation}
\Delta=\tilde\gamma^{ik}r_{,i}r_{,k}=\tilde\gamma^{00}\dot r^2-r_{,n}^2. 
\end{equation}
It follows then, that
\begin{equation}
{r_{,n}}_{\big|\Sigma}=\sigma\sqrt{\dot\rho^2-\Delta}, \quad \sigma=\pm1,
\end{equation}
where the sign function $\sigma=\pm1$ determines whether the radius $r$ increases in the outward normal direction or it decreases. Thus, its value on the either side of the shell  gives us the information about the global geometry  of the whole spherically symmetric space-time. Substituting this back into the shell equation, one gets
\begin{equation}
\sigma_{\rm in}\sqrt{\dot\rho^2-\Delta_{\rm in}}
-\sigma_{\rm out}\sqrt{\dot\rho^2-\Delta_{\rm out}}
=\frac{4\pi G}{\rho^2}S_0.
\end{equation}
The famous Birkhoff theorem states, actually, that for the vacuum spherically symmetric solution  of the Einstein equations
\begin{equation}
\Delta=-\rho^2\left(1-\frac{2Gm}{\rho}-\frac{\Lambda}{3}\rho^2\right),
\end{equation}
where $m$ is the so called Schwarzschild mass. Since we supposed that our shell is the innermost one, then, $m_{\rm in}=0$, and
\begin{eqnarray}
\Delta_{\rm in}\Big|_\Sigma &=& -\rho^2\left(1-\frac{\Lambda}{3}\rho^2 \right);  \\
\Delta_{\rm out}\Big|_\Sigma &=& -\rho^2\left(1-\frac{2Gm}{\rho}-\frac{\Lambda}{3}\rho^2 \right).
\end{eqnarray}
The value $m_{\rm in}=0$ corresponds to the de Sitter $(\Lambda>0)$, Anti-de Sitter $(\Lambda>0)$ or Minkowski $(\Lambda=0)$ space-time inside the shell. In all of these manifolds the Weyl tensor is identically zero, $C_{\mu\nu\lambda\sigma}\equiv0$. 

Squaring the above equation we get 
\begin{equation}
m=\frac{4\pi S_0}{\rho^3}\sigma_{\rm in}\sqrt{\dot\rho^2-\Delta_{\rm in}}
-\frac{8\pi^2 GS^2_0}{\rho^5},
\end{equation}
so, given the initial condition ($\rho_0$ and $\dot\rho_0$) and the energy content of the shell $(S_0)$ we determine the Schwarzschild mass $m$. Thus we conclude, that the trajectory of the traceless shell immersed in vacuum, can be completely restored (for given shell and given initial conditions), but this requires the knowledge of the vacuum solutions both inside and outside the shell. What concerns the values of the extrinsic curvature $\tilde K|_\Sigma$, their calculation is possible only if one knows the trajectory (i.\,e., $\rho(\tau)$) even in the simplest case $\tilde R=2$.

\subsection{Thin shells in Weyl gravity}

The action integral for the Weyl gravity (which is also known as conformal gravity or pure conformal gravity) is
\begin{equation}
S_{\rm tot}=-\alpha_0\int C^2\sqrt{-g}\,dx+S_{\rm m},
\end{equation}
where $C^2$ is the square of the Weyl tensor (see Section \ref{prelim} above). The equations of motion (already conformally transformed)  are the Bach equations
\begin{equation}
\hat B_{\mu\nu}=\frac{1}{8\alpha_0}\hat T_{\mu\nu}.
\end{equation}
The spherically conformal gravity was investigated in details in\cite{BDE16-BDE17}, and here we write down only the final result for the corresponding Bach equations of motion:
\begin{equation}
\hat B_{ik}=-\frac{1}{6}\left(\tilde R^{||p}_{||p}\tilde\gamma_{ik}-\tilde R_{||ik}+  
\frac{\tilde R^2-4}{4}\tilde\gamma_{ik}\right)=\frac{1}{8\alpha_0}\hat T_{ik},
\end{equation}
where ``$||$'' --- covariant derivative with respect to $\tilde \gamma_{ik}$ and 
$\hat B^\mu_\mu=0$, $B^2_2=B^3_3=-B^p_p/2$. Note, that there is no any trace of the radius $r$ in the equations of motion. Also, the energy-momentum tensor must be traceless, $\hat T^\lambda_\lambda=0$, this follows not only from the equations of motion, but from demanding the conformal invariance of the total action as well.

In the Gauss normal coordinate system the above set of equation of motion takes the form
\begin{eqnarray}
\hat B_{00}&=&\frac{1}{6}\left(\frac{\tilde R^2-4}{4}-\tilde R_{,nn}\right)\tilde\gamma_{00}
=\frac{1}{8\alpha_0}\hat T_{00};   \\
\hat B_{0n}&=&-\frac{1}{6}\left(\dot{\tilde{R}}_{,n}+\tilde K\dot{\tilde{R}}_{,n}\right)
=\frac{1}{8\alpha_0}\hat T_{0n};   \\
\hat B_{nn}&=&\frac{1}{6\tilde\gamma_{00}}\left(\ddot{\tilde{R}}
-\frac{\dot{\tilde{\gamma}}_{00}}{\tilde{\gamma}_{00}}\dot{\tilde{R}}
+\tilde{\gamma}_{00}\tilde K\tilde{R}_{,n}\right)-\frac{\tilde R^2-4}{24}
=\frac{1}{8\alpha_0}\hat T_{00}. 
\end{eqnarray}
Remember, that we supposed the absence of the double layer. This means, that the two-dimensional curvature scalar $\tilde R$ must be continuous on $\Sigma$, i.\,e., $[\tilde R]\Big|_\Sigma=0$ and 
\begin{eqnarray}
\tilde R_{,n}&=&[\tilde R_{,n}]\Theta(n)+\tilde R_{,n}^{(-)};   \\
\tilde R_{,nn}&=&[\tilde R_{,n}]\delta(n)+[\tilde R_{,nn}]\Theta(n)
+\tilde R_{,nn}^{(-)}.
\end{eqnarray}
The continuity of $\tilde R$ leads also to the continuity of both $\tilde K$ and $\tilde K_{,n}$, since
$\tilde R=-2\tilde K_{,n}+\tilde K^2$. As a result, we get the following matching conditions on the shell:
\begin{equation}
\left\{
\begin{array}{ll}	
-[\tilde R_{,n}]=\frac{3}{4\alpha_0}\hat S_{00}; &  \\
-[\tilde R_{,nn}]=\frac{3}{4\alpha_0}[\hat T_{00}]. &  
\end{array}	\right.
\end{equation}
\begin{equation}
\left\{
\begin{array}{ll}	
S_{0n}=0; &  \\
-[\dot{\tilde{R}}_{,n}]=\frac{3}{4\alpha_0}[\hat T_{0n}]. &  
\end{array}	\right.
\end{equation}
\begin{equation}
\left\{
\begin{array}{ll}	
S_{nn}=0 &  \\
-\tilde K[\tilde{R}_{,n}]=\frac{3}{4\alpha_0}[\hat T_{nn}]. &  
\end{array}	\right.
\end{equation}
Or,
\begin{equation}
\left\{
\begin{array}{ll}	
\dot{\tilde{S}}_0^0-[\hat{T}_0^n]=0; &  \\
-[\tilde R_{,n}]=\frac{3}{4\alpha_0}\hat S_0^0; &  \\
-\tilde K[\tilde S_0^0]+[\hat{T}_n^n]=0. &  
\end{array}	\right.
\end{equation}
If on both sides we have the vacuum conditions, then $\tilde K|_\Sigma=0$. 

Let us assume that inside the shell the vacuum solution corresponds to $\tilde R=2$ (this is in order to compare with the case considered above for the Einstein gravity). Then the relation $\tilde R=-2\tilde K_{,n}+\tilde K^2$, together with the boundary conditions $\tilde K(0,\tilde\tau)=1$ and $\tilde\gamma_{00}(0,\tilde\tau)=1$, allows to restore completely the conformally transformed metric inside the shell. Namely, for $n<0$ we have 
\begin{eqnarray}
\tilde K&=&-\tanh[n+\varphi(\tau)],  \\
\tilde K_{\big|\Sigma}&=&\tilde K(\tau,0)=0 \quad \Rightarrow  \quad \varphi=0
\quad \Rightarrow \quad \tilde K=-\tanh n,   \\
\tilde\gamma_{00}&=&\cosh^2\!n, \quad \Rightarrow \\ 
d\tilde s^2_2&=&\cosh^2\!n\,d\tilde\tau^2-dn^2.
\end{eqnarray}
Surely, the radius $r(n,\tilde\tau)$ remains the arbitrary function! The only thing that is definitely true is the following. If the sphere of zero radius, $r=0$, lies inside our shell, then it should be located at $n=-\infty$.

\subsection{Thin shells in Weyl$+$Einstein gravity}

The total action integral in this case is
\begin{equation}
S_{\rm tot}=-\alpha_0\int C^2\sqrt{-g}\,dx+\frac{1}{16\pi G}\int(R-2\Lambda)\sqrt{-g}\,dx+S_{\rm m}.
\end{equation}
We are not going to write down repeatedly the equations of motion and come straight to the matching conditions for the spherically symmetric shell (without the double layer):
\begin{eqnarray}
0&=&\dot{\hat{S}}^0_0-\frac{\dot r}{r}(\hat S^0_0+\hat S^2_2)+[\hat T^n_0];   \\
-2\alpha_0[\tilde R_{,n}]&=&\hat S^0_0-\hat S^2_2;   \\
-r[r_{,n}]&=&4\pi G(\hat S^0_0+2\hat S^2_2) \\
-2\alpha_0[\tilde R_{,nn}]&=&\tilde K(\hat S^0_0+\hat S^2_2)-[\hat T^0_0-\hat T^n_n-\hat T^2_2];   \\
r[r_{,nn}]&=&-\frac{4\pi G}{3}\left(\tilde K(\hat S^0_0+2\hat S^2_2)+Tr [\tilde T]\right);   \\ 
-[\Delta]&=&\frac{8\pi G}{3}(\tilde K\hat S^0_0+[\tilde T^n_n]). 
\end{eqnarray}
For the traceless shell  in the vacuum we have
\begin{eqnarray}
\dot{\hat{S}}^0_0&=&0;   \\
-[\tilde R_{,n}]&=&\frac{3}{4\alpha_0}\hat S^0_0;   \\
-r[r_{,n}]&=&0; \\
4\alpha_0[\tilde R_{,nn}]&=&\tilde K\hat S^0_0;   \\
r[r_{,nn}]&=&0;   \\ 
-[\Delta]&=&\frac{8\pi G}{3}\tilde K\hat S^0_0. 
\end{eqnarray}
Since $\Delta|_\Sigma=\dot\rho^2-r_{,n}^2$, the condition $[r_{,n}]=0$ implies $[\Delta]=0$, and from this it follows that $K|_\Sigma=0$. Thus, we are left with exactly the same shell equations as before, in the case of the pure conformal gravity. And no trace of the trajectory $\rho(\tilde\tau)$. 

But! Let us choose the same vacuum inside as before, i.\,e., with $\tilde R=2$. In the case of Weyl$+$Einstein gravity it means that inside the shell se have (Anti)-de Sitter space-time (so the Weyl tensor is zero). Thus, we are left with the spherically symmetric Einstein equations (there are four of them) for radius $r(n,\tilde\tau)$ and metric tensor $\tilde\gamma_{ik}$, $(i,k=0,1$). However, the latter we already know because of $\tilde K|_\Sigma=0$! (See the preceding Section.) Remarkably enough, the solution exists:
\begin{equation}
r=\frac{1}{\cosh n(C_1\cos\tilde\tau+C_2\sin\tilde\tau)+d_0|\sinh n|},
\end{equation}
\begin{equation}
C^2_1+C^2_2=\frac{\Lambda}{3}+d^2_0.
\end{equation}
For the trajectory, $\rho(\tilde\tau)=\rho(0,\tilde\tau)$, we obtain
\begin{equation}
\rho(\tau)=\frac{1}{C_1\cos\tilde\tau+C_2\sin\tilde\tau}=\frac{1}{C_0\sin[\tilde\tau-\tilde\tau_0]}.
\end{equation}
The constants of integration $C_1$ and $C_2$ are determined by the initial conditions $\rho(0)=\rho_0$ and  $\dot\rho(0)=\dot\rho_0$. It seems that it is possible for the shell to reach the infinity $\rho=\infty$, for finite time interval $(\tilde\tau=\tilde\tau_0)$. But, remember, that it is not the genuine proper time, the latter being defined by $d\tau=\rho(\tilde\tau)d\tilde\tau$ and 
\begin{equation}
\Delta\tau=\int^{\tilde\tau_0}\!\!\frac{d\tilde\tau}{C_0\sin[\tilde\tau-\tilde\tau_0]}=\infty 
\end{equation}
as it should be!

\section{Conclusion}

Let us now discuss briefly what we did and what we obtained. We have built the thin shell formalism for the special case of the spherically symmetric conformal gravity. We considered in more details the traceless thin shells immersed in the vacuum and compared their behavior for the three different gravitational theories, namely, General relativity (= Einstein gravity), pure conformal (Weyl) gravity and Weyl$+$Einstein gravity. The second and third ones are special cases of the so called quadratic gravity. The main difference between General Relativity and generic quadratic gravities is that the latter ones contain the fourth order derivatives of the metric tensor, while the former is confined to the second derivatives. 

It is this feature that causes quite different behavior of thin shells in these theories. In General relativity, in order to calculate the thin shell trajectory (which is just the radius as a function of the proper time of the observer, sitting on the shell), one needs to know the solutions in the bulk on both sided of the shell, but in the Weyl  (pure conformal) gravity the shell trajectory is completely undetermined. It seems quite unphysical because the motion of the innermost shell serves as a boundary condition for the outside bulk region. 

The situation in the Weyl$+$Einstein gravity even more tricky. We give an example of the vacuum solution inside the innermost (i.e., the first) shell, when the matching conditions allows to determine unambiguously (inside the shell) not only the conformally transformed metric in Gauss normal coordinates, but also to restore the radius (which is chosen as the conformal factor). And, thus, the shell trajectory itself (of course, up to the initial conditions). We are sure that the theory of the thin shell in the Weyl$+$Einstein gravity deserves further investigation.

\section*{Acknowledgments}

We acknowledge A. L. Smirnov for helpful discussions. This study was supported by the Russian Foundation for Basic Research (project no. 15-02-05038-a).


\begin{thebibliography}{00}   

\bibitem{Isr} W. Israel, {\it Il Nuovo Cim. B} {\bf 44} (1966) 1.

\bibitem{Cruz67} V. de la Cruz and W.Israel, {\it Il Nuovo Cim.} {\bf 51} (1967) 744.

\bibitem{Lake79} K. Lake, {\it Phys. Rev. D} {\bf 19} (1979).

\bibitem{BKT83} V. A. Berezin, V. A. Kuzmin and I. I. Tkachev, {\it Phys. Lett. B} {\bf 120} (1983) 91.

\bibitem{BKT83b} V. A. Berezin, V. A. Kuzmin and I. I. Tkachev, {\it Phys. Lett. B} {\bf 124} (1983) 479.

\bibitem{BKT83c} V. A. Berezin, V. A. Kuzmin and I. I. Tkachev, {\it Phys. Lett. B} {\bf 130} (1983) 23.

\bibitem{BKT84} V. A. Berezin, V. A. Kuzmin and I. I. Tkachev, {\it Sov. Phys.—JETP} {\bf 59} (1984) 459.

\bibitem{Gyulassy84} M. Gyulassy, K. Kajantie, H. Kurki-Suonio and L. McLerran. {\it Nucl. Phys. B} {\bf 237} (1984) 477. 

\bibitem{Lake84} K. Lake, {\it Phys. Rev. D} {\bf 29} 1861 (1984).

\bibitem{Grand84} T. De Grand and K.Kajantie, Phys. Lett. 147B, 273 (1984). [5] A.Aurilia, G.Denardo,

\bibitem{Legovini84} F.Legovini and E.Spallucci, Phys. Lett. 147B, 258 (1984), Nucl. Phys. B252, 523 (1985).

\bibitem{BKT85} V. A. Berezin, V. A. Kuzmin and I. I. Tkachev, {\it Pis’ma Zh. Eksp. Teor. Fiz.} {\bf 41} (1985) 446.

\bibitem{Sato86} H. Sato, {\it Progr. Theor. Phys.} {\bf 76} (1986) 1250. 

\bibitem{Laguna86} P. Laguna-Castillo and R. A. Matzner, {\it Phys. Rev. D} {\bf 34}, 2913 (1986).

\bibitem{BKT87} V. A. Berezin, V. A. Kuzmin and I. I. Tkachev, {\it Sov. Phys. —JETP} {\bf 93} (1987) 1159.

\bibitem{BKT87b} V. A. Berezin, V. A. Kuzmin and I. I. Tkachev, {\it Phys. Rev. D} {\bf 36} (1987) 2919.

\bibitem{Blau87} S. K. Blau, E. I. Guendelman and A. H. Guth {\it Phys. Rev. D} {\bf 35} (1987) 1747.

\bibitem{Farhi87} E.Farhi and A.Guth, Phys. Lett. B183, 149 (1987).

\bibitem{Aurilia89} A. Aurilia, M. Palmer and E. Spallucci, {\it Phys. Rev. D} {\bf 40} (1989) 2511.

\bibitem{Farhi90} E. Farhi, A. H. Guth and J. Guven, {\it Nucl.Phys. B} {\bf 339} (1990) 417.

\bibitem{BKT90}  V. A. Berezin, V. A. Kuzmin and I. I. Tkachev, {\it Int. J. Mod. Phys. A} {\bf 45} (1990) 4639. 

\bibitem{BKT91}  V. A. Berezin, V. A. Kuzmin and I. I. Tkachev, {\it Physica Scripta} {\bf 36} (1991) 269.

\bibitem{BKT91b} V. A. Berezin, V. A. Kuzmin and I. I. Tkachev, {\it Phys. Rev. D} {\bf 43} (1987) R3112.

\bibitem{Aguirre05} A. Aguirre and M. Johnson, {\it Phys. Rev. D} {\bf 72} (2005) 103525.

\bibitem{Lee07} B-H. Lee, W. Lee, S. Nam and C. Park, {\it Phys. Rev. D} {\bf 75} (2007) 103506.

\bibitem{Mann08} R. B. Mann and J. J. Oh, {\it Phys. Rev. D} {\bf 74} (2006) 124016; Erratum {\it Phys. Rev. D} {\bf 77} (2008) 129902.

\bibitem{Boulware73} D. G. Boulware, {\it Phys. Rev. D} {\bf 8} (1973) 2363.

\bibitem{Frolov74} V. P. Frolov, {\it Sov. Phys. JETP} {\bf 38} (1974) 393. 

\bibitem{Berezin05} V. Berezin, V. Dokuchaev, Y. Eroshenko and A. Smirnov {\it Class. Quant. Grav.} {\bf 22} (2005) 4443.

\bibitem{Dokuch07} V. I. Dokuchaev  and S. V. Chernov 2007 {\it J. Exp. Theor. Phys. Lett.} {\bf 85} (2007) 595.

\bibitem{Dokuch08} V. I. Dokuchaev  and S. V. Chernov  {\it J. Exp. Theor. Phys.} {\bf 107} (2008) 203.

\bibitem{Chernov08} S. V. Chernov and V. I. Dokuchaev {\it Class. Quant. Grav.} {\bf 25} (2008) 015004.

\bibitem{Dokuch10} V. I. Dokuchaev and S. V. Chernov,  {\it J. Exp. Theor. Phys.} {\bf 111} (2010) 570.

\bibitem{Berezin14} V. A. Berezin, V. I. Dokuchaev, arXiv:1404.2726 [gr-qc].

\bibitem{Berezin14b} V. A. Berezin, V. I. Dokuchaev, arXiv:1404.2727 [gr-qc].

\bibitem{Parker69} L. Parker, {\it Phys. Rev.} {\bf 183} (1969) 1057.

\bibitem{Grib70}  A. A. Grib and S. G. Mamaev, {\it Sov. J. Nucl. Phys.} {\bf 10} (1970) 722.

\bibitem{Zeld70} Ya. B. Zel’dovich, {\it JETP Lett.} {\bf 9} (1970) 307.

\bibitem{Zeld71} Ya. B. Zel’dovich and L. P. Pitaevsky, {\it Comm. Math. Phys.} {\bf 23} (1971) 185.

\bibitem{Zeld72} Ya. B. Zel’dovich and A. A. Starobinskii, {\it Sov. Phys. JETP} {\bf 34} (1972) 1159.

\bibitem{Parker73} L. Parker and S. A. Fulling, {\it Phys. Rev. D} {\bf 7} (1973) 2357.

\bibitem{Fulling73} S. A. Fulling, {\it Phys. Rev. D} {\bf 7} (1973) 2850.

\bibitem{Parker73b} B. L. Hu, S. A. Fulling and L. Parker, {\it Phys. Rev. D} {\bf 8} (1973) 2377.
	
\bibitem{Parker74} L. Parker and S. A. Fulling, {\it Phys. Rev. D} {\bf 9} (1974) 341.

\bibitem{Parker74b} S. A. Fulling, L. Parker and B. L. Hu, {\it Phys. Rev. D} {\bf 10} (1974) 3905.

\bibitem{Parker74c} S. A. Fulling and L. Parker, {\it Ann. Phys.} {\bf 87} (1974) 176.

\bibitem{Lukash74} V. N. Lukash and A. A. Starobinskii, {\it Sov. Phys. JETP} {\bf 39} (1974) 742.

\bibitem{Zeld77}  Ya. B. Zel’dovich and A. A. Starobinskii, {\it JETP Lett.} {\bf 26} (1977) 252.

\bibitem{Alstar} A. A. Starobinsky, {\it Phys. Lett. B}{\bf 91} (1980) 99.

\bibitem{Sakh} A. D. Sakharov, {\it Sov. phys. Doklady} {\bf 12} (1968) 1040 [in Russian: DAN (1967)].

\bibitem{Frolov} H. -H. von Borzeszkowski and V. P. Frolov, {\it Annalen der Physik} {\bf 37} (1980) 285.

\bibitem{BDE16} V. Berezin, V. Dokuchaev and Yu. Eroshenko, {\it JCAP} {\bf 01} (2016) 019.
 
\bibitem{BDE16b} V. Berezin, V. Dokuchaev and Yu. Eroshenko, {\it J. Mod. Phys. A} {\bf 31} (2016) 1641004.

\bibitem{bde17a}  V. A. Berezin, V. I. Dokuchaev, Yu. N. Eroshenko, Russian Phys. J. {\bf 59} (2017) 1819.

\bibitem{BDE17} V. Berezin, V. Dokuchaev and Yu. Eroshenko, {\it JCAP} {\bf 01} (2017) 018.

\bibitem{Senovilla15} J. M. M. Senovilla, {\it J. of Phys. Conf. Ser.} {\bf 600} (2015) 012004.

\bibitem{Senovilla17} E. F. Eiroa, G. F. Aguirre and J. M. M. Senovilla, {\it Phys. Rev. D} {\bf 95} (2017) 124021.

\bibitem{Berezin01} V. Berezin and M. Okhrimenko, {\it Class. Quant. Grav.} {\bf 18} (2001) 2195.

\end{thebibliography}
\end{document}